\documentclass{optica-article}

\journal{opticajournal} 

\articletype{Research Article}
\usepackage{wasysym}
\usepackage{siunitx} 
\usepackage{physics}
\makeatletter
\renewcommand{\@biblabel}[1]{#1. }
\renewcommand{\@dotsep}{500}
\renewcommand{\@pnumwidth}{0em}
\renewcommand{\l@figure}[2]{
\@dottedtocline{1}{1.5em}{2em}{Figure #1}{}\vspace{15pt}}

\usepackage{lineno}

\begin{document}

\title{Magnetic field free nonreciprocity in tapered atomic cladded nano waveguide }

\author{Ilan Sher\authormark{1,$\dagger$}, Benyamin Shnirman\authormark{2,3,$\dagger$}, Arieh Grosman\authormark{1}, Roy Zektzer \authormark{4},Markus Greul\authormark{3}, Mathias Kaschel\authormark{3}, Tilman Pfau\authormark{2}, Robert Löw\authormark{2},  and Uriel Levy\authormark{1,*}}

\address{
\authormark{1}Institute of Applied Physics, The faculty of science, The Center for Nanoscience and Nanotechnology, The Hebrew University of Jerusalem, Jerusalem, 91904, Israel\\
\authormark{2}5th Institute of Physics, Center for Integrated Quantum Science and Technology (IQST), University of Stuttgart, Pfaffenwaldring 57, 70569 Stuttgart, Germany\\
\authormark{3}Institut für Mikroelektronik Stuttgart (IMS CHIPS), Allmandring 30a, 70569 Stuttgart, Germany\\
\authormark{4}Faculty of Engineering, Bar-Ilan University, Ramat-Gan 5290002, Israel\\
\authormark{$\dagger$}These authors contributed equally\\
}

\email{\authormark{*}ulevy@mail.huji.ac.il} 

\begin{abstract*} 

Optical nonreciprocity is a fundamental requirement for modern optical communications and quantum information processing, where it is essential to protect sensitive sources from destabilizing feedback and preserving quantum coherence. Conventional nonreciprocal devices are based on the Faraday effect; however, their dependence on bulky permanent magnets poses a significant barrier to chip-scale integration and scalability. Moreover, the application of a magnetic field is undesired in many quantum atomic systems. In this work, we demonstrate magnet-free optical nonreciprocity on a fully integrated platform utilizing a Nanophotonic Alkali Silicon Waveguide (NASWAG) interfaced with hot rubidium vapor. By employing velocity-selective optical pumping (VSOP), we break time reversal symmetry by taking advantage of the Doppler effect-generated by the thermally moving atoms, a phenomenon traditionally viewed as a limitation in atomic spectroscopy. We show that the use of suspended tapered waveguides significantly mitigates transit-time broadening, thereby enabling the observation of a robust nonreciprocal response. We further characterize the dependence of the isolation contrast on pump power, finding that the experimental measurements and numerical simulations correspond and provide mutual support for the underlying physical model. With proper optimization, the demonstrated effect may be used in the future for applications such as magnetic free optical isolators.

\vspace{1cm}
\end{abstract*}
\section{Introduction}
\noindent

Optical nonreciprocity is frequently illustrated by the analogy of a one-way mirror. However, this comparison is physically imprecise, as asymmetric transmission alone does not satisfy the rigorous criteria for nonreciprocity. A classic counter-example is the junction between a single-mode and a multimode waveguide: while light transmits efficiently from the single-mode to the multimode section, the reverse process-though suffering from modal mismatch-still permits the fundamental mode to propagate back to the single-mode side. True nonreciprocity requires more than mere spatial asymmetry; a device must block all back-propagating light regardless of its modal properties. Formally, this requires a violation of the Lorentz reciprocity theorem, which dictates that the electromagnetic response of a linear reciprocal system remains symmetric under the interchange of source and detector \cite{Jalas2013, Caloz2018ElectromagneticNonreciprocity, Shi2015LimitationsReciprocity,Barzanjeh2025}.

To overcome this theorem in practical applications, the most established approach involves using the magneto-optic Faraday effect to break time-reversal symmetry. Consequently, commercial optical isolators typically employ magneto-optical materials, such as yttrium iron garnet (YIG) or terbium gallium garnet (TGG), subjected to a biased magnetic field and positioned between polarizers offset by 45$^\circ$ \cite{Saleh2019}. While these devices offer high isolation ratios (>90 dB) and high bandwidths, their reliance on bulky permanent magnets imposes a fundamental limit on scalability. The resulting footprint and potential for magnetic crosstalk render them incompatible with the dense integration required for modern photonic integrated circuits (PICs). Furthermore, they require the application of a magnetic field, which may be undesirable for many practical applications. 

This scalability bottleneck is particularly relevant given the rapid maturation of silicon photonics. Since the early 1990s, the field has evolved into a robust platform that mirrors the semiconductor revolution in electronics. By leveraging advanced nanolithography, silicon photonics enables the production of ultra-compact components-such as low-loss waveguides \cite{Lee2012b}, high-Q resonators \cite{Naiman2015Ultrahigh-QPlatform}, and high-speed modulators \cite{Wang2018NanophotonicModulators}. 

Parallel to these advances in solid-state photonics, alkali atoms have long served as the fundamental building blocks for quantum technology, precision metrology, and frequency stabilization. While these applications traditionally rely on macroscopic vapor cells, there is a vigorous ongoing effort to miniaturize these atomic systems onto the micro-scale by integrating alkali vapors with photonic circuits \cite{Stern2013, Stern2017, Ritter2015}. This hybrid approach provides access to cavity QED \cite{Srinivasan2024}, precision magnetic sensing \cite{Levi2025RemoteFields}, and other metrological standards \cite{Kitching2018} within a chip-scale footprint. By combining tight optical confinement with the strong third-order nonlinear susceptibility $\chi^{(3)}$ of alkali atoms, vapor-integrated photonics offers a sophisticated pathway for nonlinear optical control. However, while this plethora of devices ranging from classical telecommunication units to advanced atomic quantum sensors scales in complexity, they all face a common, critical limitation. A high-performance, magnet-free nonreciprocal component remains the "missing link" in the integrated toolkit. Robust, on-chip optical isolators are essential to protect delicate laser sources from destabilizing feedback, preserve qubit coherence, and maintain the high fidelity required for quantum measurements, ensuring stable operation of both classical and quantum photonic systems\cite{Yu2009OpticalTransitions,Ruesink2016NonreciprocityInteractions,Talker2020a}. To address the integration challenges associated with traditional Faraday rotation, several alternative mechanisms for nonreciprocity have been developed. These include spatiotemporal and dynamic modulation of the refractive index to break time-reversal symmetry \cite{Yu2009OpticalTransitions, Lira2012ElectricallyChip, Sohn2018Time-reversalCircuits, Shah1978Visible-telecomNiobate, Fang2012PhotonicModulation, Tzuang2014Non-reciprocalLight, Kittlaus2021ElectricallyPhotonics}. Another approach involves nonlinear propagation in Kerr-active media \cite{Sounas2017Non-reciprocalModulation, DelBino2018MicroresonatorEffect}. Furthermore, nonreciprocity has been demonstrated through stimulated Brillouin scattering \cite{Kim2018, Dong2015} and optomechanical interactions within various resonator systems \cite{Shen2016ExperimentalNon-reciprocity, Wanjura2023NatureSymmetry, Hafezi2012OptomechanicallyResonators, Li2017OpticalSystem, Ruesink2016NonreciprocityInteractions}. More recently, techniques leveraging chiral light-matter interactions in nanophotonic structures \cite{Sayrin2015NanophotonicAtoms,Scheucher2016QuantumAtom, Zektzer2019} and the random thermal motion of atoms \cite{Zhang2018, Hu2019,Song2021OpticalAtoms, Zhan2025OpticalShifts,Dong2021All-opticalTemperature,Song2022} have also been explored.

\begin{figure}[!htbp]
\centering\includegraphics[width=1\linewidth]{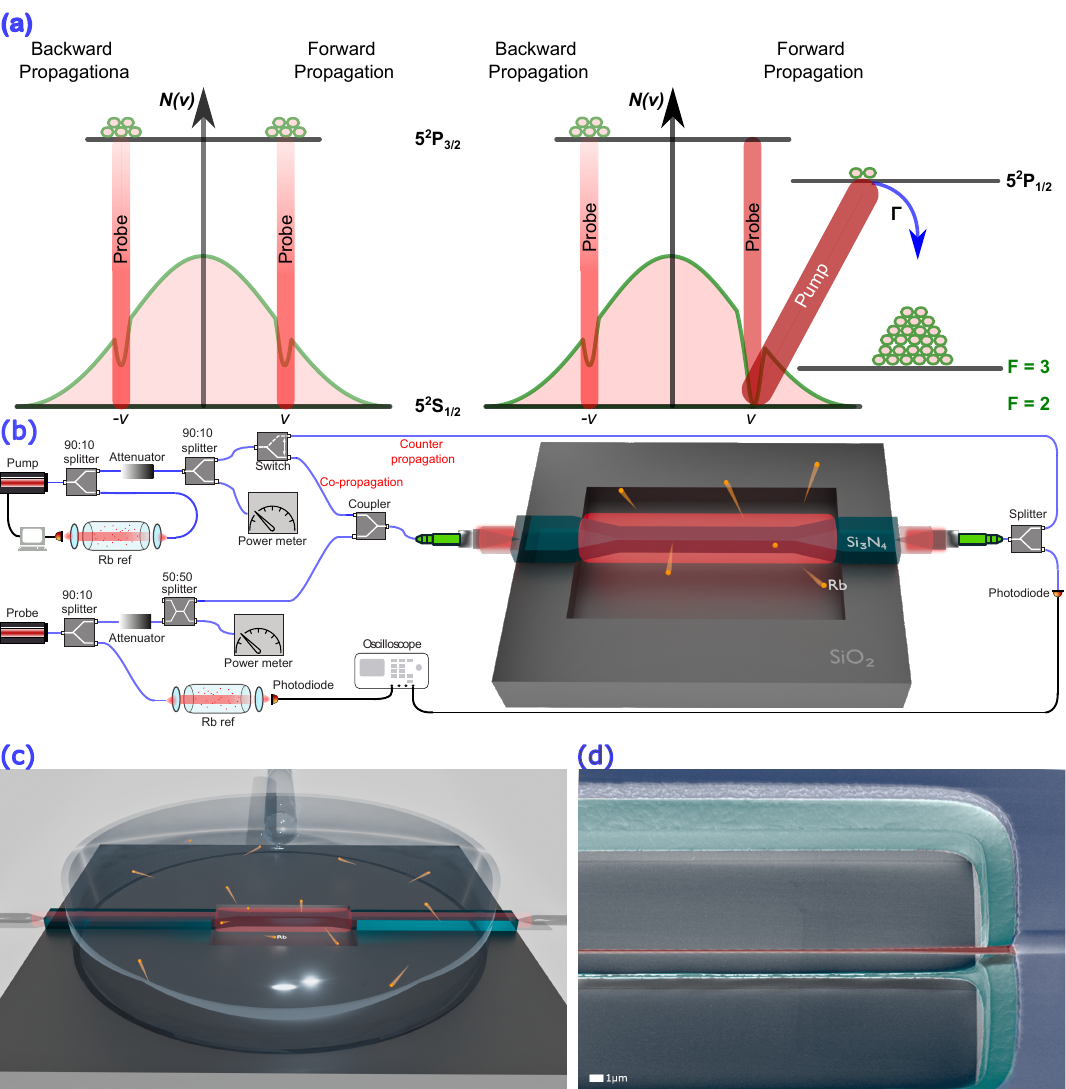}
\caption{
\textbf{(a) Mechanism of breaking reciprocity.} Atomic level scheme for Rubidium ($5^2S_{1/2}$, $5^2P_{1/2}$, $5^2P_{3/2}$) and Maxwell-Boltzmann velocity distributions (only the $F=3$ population is shown). \textbf{Left (Reciprocal case):} Weak probe beams (thin red bars) sample a symmetric velocity distribution, resulting in identical forward and backward absorption. \textbf{Right (Nonreciprocal case):} An intense pump laser (wide red bar) selectively depletes a specific velocity group via spontaneous decay $\Gamma$ into the $F=2$ state. Consequently, the co-propagating probe beam (thin red bar) experiences transparency due to localized population depletion, while the counter-propagating beam remains absorbed.
\textbf{(b) Experimental configuration.} Schematic of the setup where the $780$ nm probe and $795$ nm pump lasers are coupled to the chip via lensed fibers. Co- and counter-propagating schemes are realized by alternating the pump injection direction. The light-atom interaction occurs within the tapered waveguide region. The transmitted signal is filtered (BPF) to suppress the pump light and recorded by a photodiode.
\textbf{(c) Integrated device rendering.} 3D visualization of the nanophotonic chip featuring a tapered $\text{Si}_3\text{N}_4$ waveguide (red) encapsulated within a glass cell containing thermal rubidium (Rb) vapor.
\textbf{(d) Tapered waveguide morphology.} False-color SEM image of the suspended $\text{Si}_3\text{N}_4$ waveguide (red), tapered to a width of $\sim 150$ nm. The buried oxide (green) and oxide cladding (blue) were removed via BHF etching to expose the waveguide core. Scale bar: \SI{1}{\micro\metre}.}
\label{Fig1}
\end{figure}

In this work, we break the time reversal symmetry and demonstrate chip scale nonreciprocity without the need for a magnetic field. This is achieved by leveraging the velocity-selective optical pumping (VSOP) technique in hot rubidium vapor\cite{Hu2019,Sher2025VelocityVapor}. While the Doppler effect is considered unfavorable in most applications as it generates undesired line broadening and reduction of coherence times, here it plays a vital role in breaking Lorentz reciprocity. Specifically, a weak $\SI{780}{\nano\metre}$ probe scans over the rubidium $\ket{5^2S_{1/2},F=3}\rightarrow\ket{5^2P_{3/2},F=2,3,4}$ $D_2$ manifold, while a $\SI{795}{\nano\metre}$ pump is locked with a slight offset to the $\ket{5^2S_{1/2},F=3}\rightarrow\ket{5^2P_{1/2},F=2,3}$ transition. In the absence of the pump, the forward-propagating beam resonates with atoms moving at velocity $v_{fw}=\delta_s/k_s$, while the backward beam interacts with atoms at $v_{bw}=\delta_s/(-k_s)=-v_{fw}$. Since the atomic velocities follow a Maxwell-Boltzmann distribution, there are, on average, equal numbers of atoms with velocities $v$ and $-v$. Consequently, absorption is identical for both directions, and the system remains reciprocal. However, when a strong pump with a slight offset is applied, it depopulates the ground level for atoms with a specific velocity $v_p=\delta_p/k_p$ by continuously pumping them out of the $\ket{5^2S_{1/2},F=3}$ level via the natural decay rate $\Gamma$. Hence, atoms with $v_{fw}=v_p$ are effectively removed from the interaction, and the forward signal beam can no longer be absorbed (transparency). In contrast, the counter-propagating signal remains resonant with atoms moving at $v_{bw}=-v_p$, which are unaffected by the pump, leading to absorption. Thus, nonreciprocity is achieved through an asymmetric redistribution of ground-state atoms across different velocity classes (See Fig.~\ref{Fig1}(a)). Furthermore, due to the atomic multilevel structure, nonreciprocal behavior can be achieved for all wavelengths resonant with various atomic levels.

While the Doppler resonance shift is the core mechanism for symmetry breaking, it is always accompanied by Doppler broadening, which can obscure the spectral shift of the absorption dip between co- and counter-propagating beams. Due to the guided mode configuration, Doppler broadening is enhanced as the photon wavenumber is multiplied by the effective refractive index of the optical mode. Additionally,  in photonic waveguides, the linewidth is further impaired by large transit-time broadening, stemming from the short interaction time of flying atoms traversing the evanescent field. To overcome these limitations, we utilize the recently introduced Nanoscale Atomic Suspended Waveguide (NASWAG) \cite{Zektzer2021c, Sher2026}, which offers a significant reduction in Doppler and transit-time broadening compared to standard waveguide-on-insulator geometries. Because the optical mode extends significantly out of the NASWAG core into a space filled with rubidium vapor, thermally moving atoms encounter the light field for a longer period of time, thereby reducing transit-time broadening. Additionally, as more of the optical mode resides outside the waveguide core, its effective refractive index approaches the unity refractive index of the air. This results in a significant reduction in Doppler broadening, which depends on both the photon momentum and the atomic velocities.

\section{ Experiment}

The experimental setup is illustrated in Fig.~\ref{Fig1}(b). The 780 nm probe and 795 nm pump beams were generated by two fiber-coupled external cavity diode lasers (ECDL, Toptica™) with linewidths $<100$ kHz. The configuration allows for switching between co- and counter-propagation geometries by reversing the pump direction while keeping the probe path fixed. A reference free-space Rb cell is used to monitor the probe absorption on the $D_2$ line, while a second cell serves to lock the pump laser to the $D_1 \left(\ket{5^2S_{1/2}, F=2} \rightarrow \ket{5^2P_{1/2}}\right)$transition with an adjustable frequency offset. This detuning is experimentally optimized to $\approx 200$ MHz to balance the trade-off between spectral separation (targeting specific velocity classes) and signal amplitude, which is limited by the Maxwell-Boltzmann distribution at high velocities. Light is coupled into the Nanophotonic Alkali Silicon Waveguide (NASWAG) chip via lensed optical fibers. The output probe signal is spectrally filtered to suppress residual pump light before detection by a high-sensitivity femtowatt photoreceiver (Newport™). Power levels are regulated via variable optical attenuators and monitored in real-time to prevent saturation broadening. Precise thermal management is essential for stable operation: while the rubidium reservoir is heated to achieve the required vapor density, the chip and its glass cap are maintained at a slightly higher temperature ($T_{chip} > T_{reservoir}$). This deliberate thermal gradient prevents atomic condensation on the waveguides and optical windows, thereby improving the durability of the device.

Device fabrication was carried out on a 150 mm silicon wafer featuring a \SI{3}{\micro\meter} thick buried oxide (BOX) layer and a 250 nm silicon nitride ($\mathrm{Si_{3}N_{4}}$) device layer. The waveguide patterns were defined via electron-beam lithography and transferred into the $\mathrm{Si_{3}N_{4}}$ layer using reactive ion etching (RIE). To form the suspended waveguide-vapor interface, the cladding $\mathrm{SiO_{2}}$ was undercut using a combination of RIE and wet etching. The chip was then sealed with a glass cap using vacuum-compatible epoxy, filled with natural rubidium via distillation, and hermetically pinched off. The light-Rb interaction segments consist of \SI{150}{\nano\meter} wide, \SI{80}{\micro\meter} long tapered waveguides. Four such segments were concatenated to extend the total interaction length to \SI{320}{\micro\meter}, enhancing the absorption contrast to approximately \SI{60}{\%}.

\begin{figure}[!htbp]
\centering\includegraphics[width=1\linewidth]{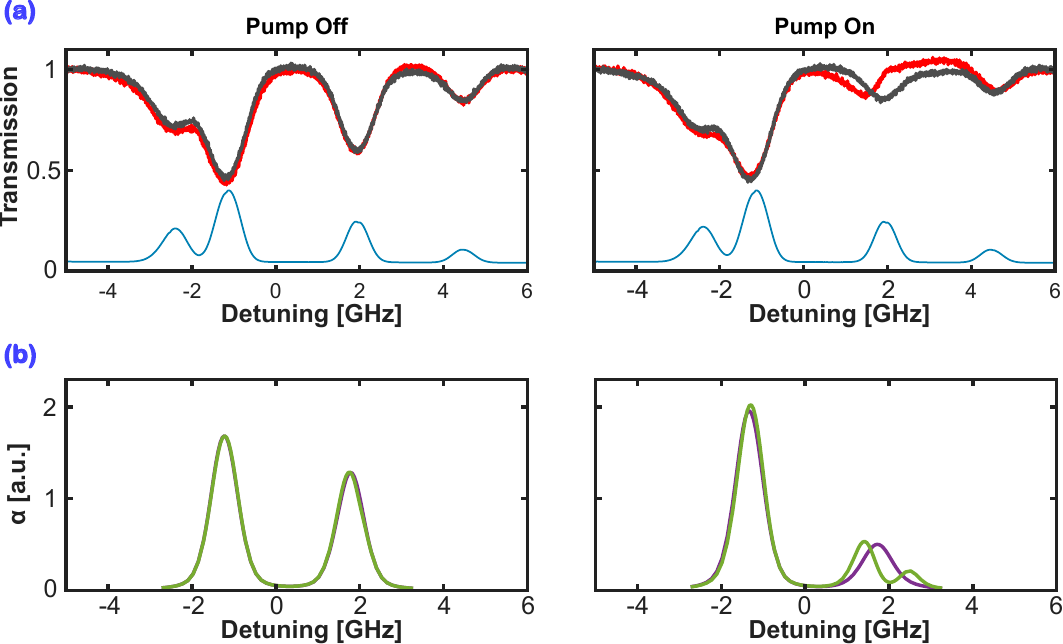}
\caption{ \textbf{(a)} Measured transmission spectra of the probe beam across the natural rubidium $D_2$ transition for co-propagating (\textit{red}) and counter-propagating (\textit{gray}) geometries. The spectra are shown for both the pump-off (\textit{left}) and pump-on (\textit{right}) configurations. An inverted absorption profile from a reference free-space vapor cell is provided in \textit{blue} for frequency calibration. \textbf{(b)} Numerically simulated absorption coefficient ($\alpha$) for the co-propagating (\textit{purple}) and counter-propagating (\textit{green}) cases, corresponding to the experimental configurations and pump conditions shown in (a).}
\label{Fig2}
\end{figure} 

\section{Results}

\begin{figure}[!htbp]
\centering\includegraphics[width=1\linewidth]{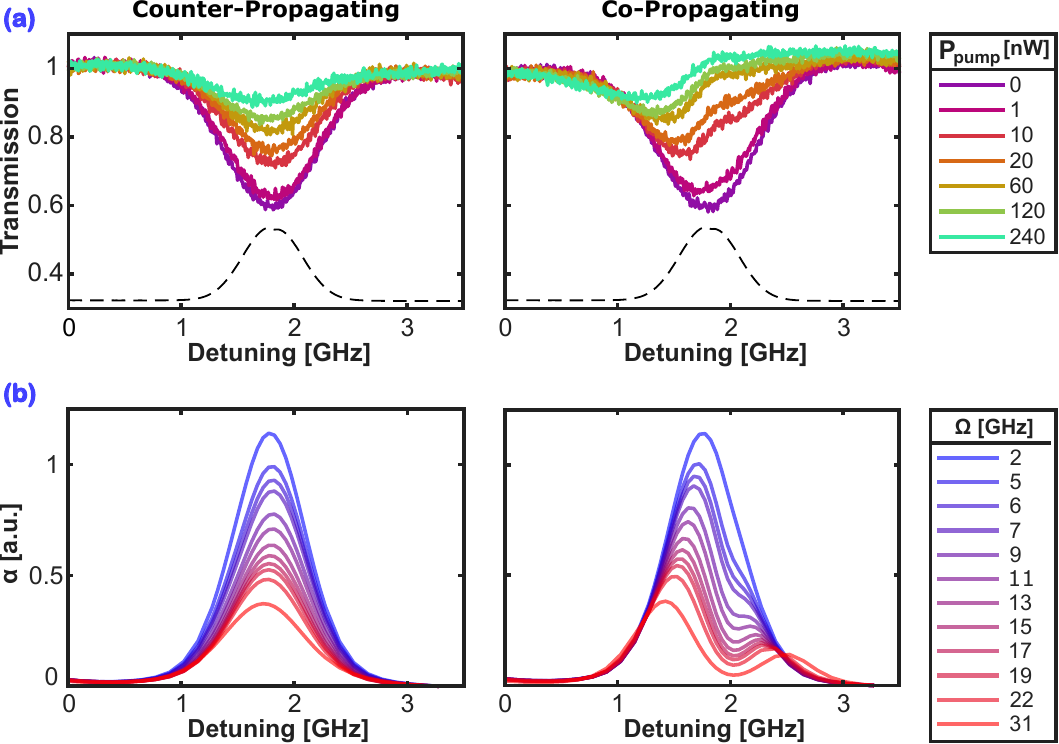}
\caption{Pump Power Dependence of Optical Nonreciprocity. \textbf{(a)} Measured transmission spectra across the Rb $F=2$ ground-state transition as a function of pump power, showing counter-propagating (\textit{left}) and co-propagating (\textit{right}) configurations. The inverted absorption profile from a commercial free-space Rb vapor cell (black dashed line) is provided as a frequency reference. \textbf{(b)} Numerically simulated absorption coefficient $\alpha$ for a range of pump powers, parameterized by Rabi frequency ($\Omega$), in the corresponding counter-propagating (\textit{left}) and co-propagating (\textit{right}) geometries, matching the experimental conditions shown in (a).}
\label{Fig3}
\end{figure}

Fig.~\ref{Fig2}(a) compares the co- and counter-propagating transmission spectra within the natural Rb $D_{2}$ manifold. In the absence of a pump beam, the transmission profiles for both directions are identical, as expected for a reciprocal system. Activating the pump beam, detuned by $\approx 200$~MHz from the $\ket{5^2S_{1/2}, F=2} \rightarrow \ket{5^2P_{1/2}}$ transition, induces a direction-dependent absorption difference for the respective hyperfine ground states of $^{85}\mathrm{Rb}$. This level of detuning was experimentally optimized to achieve the ideal balance between nonreciprocal contrast and the velocity-selective optical pumping (VSOP) induced shift of the corresponding signal. While a larger pump detuning increases the spectral separation between propagation directions - thereby facilitating isolation even in the presence of significant broadening - it simultaneously reduces the number of interacting atoms due to the exponential decay of the atomic population predicted by the Maxwell-Boltzmann distribution. This balance is highly sensitive to the total linewidth of the absorption transition, as excessive broadening can obscure the subtle nonreciprocal effect. Maintaining a narrow linewidth was a primary motivator for the development of the NASWAG architecture, which is specifically tailored to mitigate both Doppler and transit-time broadening compared to conventional Atomic Clad Waveguides (ACWGs). As schematically shown in Fig.~\ref{Fig1}(a), the resulting population transfer leads to reduced absorption in the $^{85}\mathrm{Rb}$ $F=2$ state. The characteristic timescale for hyperfine pumping on the $D_{1}$ line is on the order of $\sim \SI{1}{\micro\second}$, making the interaction time a critical parameter. In a previous ACWG design, where the evanescent tail extends only $\sim 100$~nm away of the waveguide core, a thermal atom traveling at an average speed of $\sim 300$~m/s remains within the optical mode for only a fraction of the time required for significant pumping. This is consistent with findings by Jones \textit{et al.} \cite{Jones2014SaturationVapor}, who observed negligible optical pumping in tapered optical fibers (TOFs) with a $320$~nm diameter due to high transit-time broadening. In contrast, Zektzer \textit{et al.} \cite{Zektzer2021c} reported population transfer in a tapered waveguide platform, while not being able to observe a significant population change in a standard ACWG.
In our case, one can see a broadening of the $\ket{5^{2}S_{1/2}, F=3}$ level rather than increased absorption as the optical pumping mechanism suggests. This behavior points to the mixing of atomic populations between the two ground states and can be attributed to an additional homogeneous broadening originating from unwanted contamination due to the $^{85}\text{Rb}$ distillation or due to the epoxy outgassing within the fabricated alkali-vapor photonic chip. To support these experimental findings with the physical model, we executed numerical simulations with added different values of homogeneous broadening until the best fit to the measured data was achieved. Fig.~\ref{Fig2}(b) shows a nearly identical pattern to the measured data with $125~\text{MHz}$ of homogeneous broadening added.
The dependence of the VSOP-induced shift on pump laser power is presented in Fig.~\ref{Fig3}(a). In the co-propagating configuration, increasing the pump power leads to a measurable increase of the spectral deviation from the absorption line center, accompanied by a reduction in the absorption dip depth due to enhanced optical pumping efficiency. Interestingly, the counter-propagating signal also exhibits a significant reduction in absorption despite the pump being detuned by $\approx 200$~MHz. This interaction is somewhat non-intuitive: since the pump is detuned, it primarily addresses a specific non-zero velocity class ($+v$), while the counter-propagating probe "sees" the opposite velocity class ($-v$). One might expect the counter-propagating signal to remain largely unaffected; however, the data shows a clear symmetric reduction in the absorption dip for the $^{85}\mathrm{Rb}$ $F=2$ level without a corresponding spectral shift. This effect is attributed to power broadening of the pump transition. As the pump intensity increases, its spectral "tail" eventually reaches the $v=0$ atomic velocity class. Because $v=0$ atoms are inherently insensitive to the relative propagation direction of the beams, they provide a shared interaction channel that allows the pump to deplete the ground state population for both probe directions simultaneously. This physical mechanism is well-replicated by the numerical simulations of the absorption coefficient as a function of Rabi frequency ($\Omega$), as shown in Fig.~\ref{Fig3}(b). It should be noted that the pump power within the NASWAG is not spatially constant due to longitudinal decay resulted by the Rb absorption as well as by scattering at the interfaces between the cladded and tapered segments. Consequently, the power levels indicated in Fig.~\ref{Fig3}(a) represent an averaged intensity within the tapered interaction regions. Furthermore, this non-uniformity in intensity, combined with a varying pump-to-probe power ratio along the length of the taper, results in inhomogeneous optical pumping. Such unavoidable spatial variations can introduce additional power broadening and create localized regions of under- or over-pumping, adding further complexity to the observed nonreciprocal contrast.

\section{Summary}
In conclusion, we have demonstrated velocity-selective optical pumping (VSOP) induced nonreciprocity in a hot rubidium vapor integrated with suspended nanophotonic waveguides without the application of a magnetic field. By utilizing tapered geometries to reduce both Doppler and transit-time broadening, we were able to observe a nonreciprocal response-an effect hindered by the substantial line broadening characteristic of conventional atomic-clad platforms. Using a $780$~nm probe and a $795$~nm pump, we characterized the nonreciprocal behavior as a function of pump power, finding that the experimental measurements and numerical simulations correspond and provide mutual support for the underlying physical model.

Future advancements in nano-tapered suspended waveguide designs, aimed at further suppressing line broadening and enhancing signal contrast, hold promise for the development of high-performance, magnet-free on-chip isolators. Such components are essential for the realization of robust and scalable integrated photonic platforms, supporting applications in quantum information processing, optical communications, and emerging technologies such as wafer-scale bonded atomic vapor cells \cite{Grosman2025Wafer-scaleCells,Ou2025300References,Shrestha2026EnablingCircuit}.

\begin{backmatter}
\bmsection{Funding}
Benyamin Shnirman acknowledges funding from the Zeiss Internationalization Alliance program and from the Carl-Zeiss-Stiftung (CZS Center QPhoton).

\bmsection{Acknowledgments}

\bmsection{Disclosures}

\noindent The authors declare no conflicts of interest.

\bmsection{Data Availability Statement} Data underlying the results presented in this paper are not publicly available at this time but may be obtained from the authors upon reasonable request.

\bigskip

\bmsection{Supplemental document}

\end{backmatter}

\section{References}
\label{sec:refs}

\bibliography{references}

\end{document}